\documentclass[12pt]{article}
\usepackage{amssymb,amsmath,epsfig,cite}
 \allowdisplaybreaks
\begin{document}
\title{\bf Dynamical Instability of Spherical Anisotropic Sources in $f(R,T,R_{\mu\nu}T^{\mu\nu})$ Gravity}
\author{Z. Yousaf$^1$\thanks{zeeshan.math@pu.edu.pk}, Kazuharu Bamba$^2$\thanks{bamba@sss.fukushima-u.ac.jp}, M. Z. Bhatti$^1$\thanks{mzaeem.math@pu.edu.pk} and U. Farwa$^1$\thanks{ume.farwa514@gmail.com}\\
$^{1}$ Department of Mathematics, University of the Punjab,\\
Quaid-i-Azam Campus, Lahore-54590, Pakistan\\
$^2$ Division of Human Support System,\\ Faculty of Symbiotic Systems Science,\\ Fukushima University, Fukushima 960-1296, Japan}

\date{}

\maketitle
\begin{abstract}
In this paper, we study the effects of modification of gravity on
the problem of dynamical instability of the spherical relativistic
anisotropic interiors. We have considered non-zero influence of
expansion scalar throughout during the evolutionary phases of
spherical geometry that led to the use of fluid stiffness parameter.
The modified hydrostatic equation for the stellar anisotropic matter
distributions is constructed and then solved by using radial
perturbation scheme. Such a differential equation can be further
used to obtain instability constraints at both weak field and
post-Newtonian approximations after considering a particular
Harrison-Wheeler equation of state. This approach allows us to deal
with the effects of usual and effective matter variables on the
stability exotic stellar of self-gravitating structures.
\end{abstract}

\section{Introduction}

General relativity (GR) is regarded as the foundations of relativistic astrophysics and
cosmology. The observational outcomes coming from some cosmic models like $\Lambda$-cold dark
matter turn out to consistent with various cosmological
issues besides some discrepancies such as fine-tuning and cosmic coincidence \cite{zs1}.
The accelerated cosmic expansion is strongly evidenced from the surveys of cosmic microwave background
radiations, redshift, Supernovae Type Ia and large scale structures \cite{zs2}.
These observations claimed the role of some mysterious force (dubbed as dark energy (DE))
behind this expansion. Several mathematical models have been introduced to
modify GR with the aim to explore DE and dark matter (DM). Qadir et al.
\cite{zs3} proposed that modification of Einstein gravity could be considered as a workable
toy model for various cosmological issues, like quantum gravity and DM problem.

The models of modified gravity came into their existence by modifying
the geometric portion of the Einstein-Hilbert (EH) action (for further reviews on DE and modified gravity, see, for
instance,~\cite{ya3, v41, b2b, b2a, R-DE-MG})). Nojiri and Odintsov \cite{no1}, in this direction, presented
some $f(R)$ gravity that are theoretically well-consistent for the case of our
accelerating universe. Modified gravity theories includes $f(R)$ \cite{z1fr},
Einstein-$\Lambda$ \cite{kkk1}, $f(R,T)$ \cite{z2frt}, $f(G)$ \cite{g11} gravity (here $R,~T$ and
$G$ represent Ricci scalar, trace of stress energy tensor and
Gauss-Bonnet Invariant, respectively) and $f(R,T,Q)$ \cite{z3frtrmn,z3frtrmn1} theories (where
$Q=R_{\lambda\sigma} T^{\lambda\sigma}$) etc. which
includes non-minimal coupling relating matter and geometry.

The isotropic and anisotropic nature of fluid configurations have
utmost relevance in the evolution and formation of compact
stars. Initially, Chandrasekhar \cite{6} investigated the dynamical
instability conditions of an oscillating perfect spherically symmetric celestial object.
Later on, this problem has been probed under various complicated backgrounds of
relativistic matter and geometry. Herrera et al. \cite{ck1} examined this problem by considering non-adiabatic nature of relativistic spherical
structures with weak field approximations. Chan et al. \cite{ck2} and Chan \cite{ck3} analyzed the role of
locally anisotropcity as well as heat radiation in the formulation of dynamical instability constraints of the
shearing viscous spherical matter content at both Newtonian and post-Newtonian (pN) eras. The dynamical instability of expansion-free locally anisotropic spherical stars have been
analyzed by Herrera at el. \cite{8} through perturbation scheme.

Odintsov and S\'{a}ez-G\'{o}mes \cite{z3frtrmn} studied various cosmological aspects in $f(R,T,Q)$ gravity through
construction of viable cosmological models.
Haghani et al. \cite{10} evaluated $f(R,T,Q)$ equations of motion for the case of massive
test particles by using the method of Lagrange multiplier. After assuming special case of conservation of the stress-energy tensor in this theory, they evaluated a the corresponding class of field equations. Ayuso et al. \cite{11} obtained some consistent results from the nonminimally coupled $f(R,T,Q)$ gravity and claimed that such theoretical models could be helpful to remove ceratin types of pathologies related
with the equations of higher order class of theories. Elizalde and Vacaru \cite{d11} suggested that some mathematical formulations of $f(R,T,Q)$ gravity along with their non diagonal non-holonomic equivalents could produce captivating connections between
viable quantum gravity theories. Recently, Yousaf et al. \cite{far1} analyzed the impact of particular $f(R,T,Q)$ models on the evolutionary phases of collapsing relativistic systems.

Dynamical stability is the characteristic of an object or system to retain its stable position, whenever it is disturbed due to fluctuations.
The instability/stability of celestial bodies has been discussed not only in GR but also in modified gravitational theories.. In order to analyze the dynamics of massive objects, one can use N and pN approximations in the corresponding hydrodynamical equation. Gravitational collapse (GC) of massive or dense stars and stability/instability investigation of massive
objects have attained great interest of researchers in relativistic
astronomy \cite{gc1,gc2,gc3,gc4,gc5,gc6,gc6a,gc7,gc8,gc9,gc10,gc11,gc12,gc13,gc14,gc15,gc16}. Adhav \cite{ya12} and Sahoo et al. \cite{sahoo1} evaluated some
exact analytical solution by assuming a particular class $f(R,T)$ models. The discussion of inhomogeneous energy density
in the regie of Einstein theory has also been discussed widely by Herrera et al. \cite{herr1}.

Capozziello et al. \cite{zs25} studied GC of a dust cloud through dispersion relations and perturbation
scheme and calculated certain unstable limits for the onset of collapsing phenomenon.
Cembranos et al. \cite{zs26} analyzed GC of an inhomogeneous stellar objects in order to check
large scale structure formation with the help different early time
$f(R)$ models. Ceratin modified gravity models are likely to host
supermassive structures with comparatively smaller radii as that in \cite{zs31}.
Yousaf et al. \cite{13a} inferred that dark source terms induced from some models of $f(R,T)$ gravity
could be treated as effective tools to study collapse of non-interacting particles. Baffou et
al. \cite{13b} performed stability analysis with the help of
de-Sitter and power law solution in $f(R,T,Q)$ gravity and found
that extra curvature $f(R, T,Q)$ terms could help to understand
early evolutionary cosmic stages.

The aim of this paper is to check the role of $f(R,T,Q)$ gravity on the stability of self-gravitating celestial body
The format of paper is as follows. We present the fundamental formalism to form
the corresponding field equations in section \textbf{2}. Section
\textbf{3} is devoted to formulate static as well as non-static perturbed field as well as
conservation laws. After using particular choice od equation of state, we have explored hydrodynamical equation
and instability constraints with both N and pN approximations. In the last section, we conclude our main findings.

\section{$f(R,T,R_{\mu\nu}T^{\mu\nu})$ Theory of Gravity and Field Equations}

The theories with generic functions of the form $f(R,T,R_{\mu\nu}T^{\mu\nu})$ in the action have
attracted the attention of several theoretical astrophysicists. Odintsov
and S\'{a}ez-G\'{o}mez \cite{z3frtrmn} claimed that such theories could provide some useful insights
provided by Ho\v{r}ava-like gravity under some conditions. Thus, such theories
could be considered as a theoretical bridge between Ho\v{r}ava-Lifshitz gravity and
modified theories of gravity. This section is devoted to discuss some basic
formulations of $f(R,T,R_{\mu\nu}T^{\mu\nu})$ theory. We will also formulate
the corresponding field equations for spherical anisotropic self-gravitating systems.

\subsection{Basic Formalisms of $f(R,T,R_{\mu\nu}T^{\mu\nu})$ Equations}

The formulation of  $f(R,T,R_{\gamma\delta}T^{\gamma\delta})$
gravity is based on the contribution of strong association between
geometry and matter where the Ricci scalar in usual EH
action, is replaced with generic function of $R,~T$ and
$R_{\gamma\delta}T^{\gamma\delta}$. The modified
action can be written as
\begin{equation}\label{1}
I_{f(R,T,Q)}=\frac{1} {2}\int d^4x\sqrt{-g} [f(R,T,Q)+
\textit{L}_m],
\end{equation}
where, $\textit{L}_m$ is the relative matter Lagrangian density. The expression for energy-momentum tensor is
given by the following relation
\begin{align}\label{2}
&T_{\lambda\sigma}^{(m)}=-\frac{2}{\sqrt{-g}}\frac{\delta(\sqrt{-g}\textit{L}_m)}
{\delta{g^{\lambda\sigma}}}.
\end{align}
After applying variations in the above equation with respect to $g_{\lambda\sigma}$, we get
\begin{align}\nonumber
&-G_{\lambda\sigma}(f_{Q}\textit{L}_m - f_R) - g_{\lambda\sigma}
\left\{\frac{f} {2}-\Box f_R
-\frac{R}{2}f_R-\frac{1}{2}\nabla_\pi\nabla_\rho(f_{Q}T^{\pi\rho})
-\textit{L}_mf_T\right\}\\\nonumber
&+2f_QR_{\pi(\lambda}T_{\sigma)}^{~\pi}+\frac{1}{2}\Box(f_QT_{\lambda\sigma})-\nabla_\pi\nabla_{(\lambda}
[T^\pi_{~\sigma)}f_Q]-2\left(f_Tg^{\pi\rho}+f_QR^{\pi\rho}\right)\frac{\partial^2\textit{L}_m}{\partial
g^{\lambda\sigma}\partial g^{\pi\rho}}\\\label{3} &-
T_{\lambda\sigma}^{(m)}(f_T+\frac{R}{2}f_Q+1)-\nabla_\lambda\nabla_\sigma f_R=0,
\end{align}
where $~\nabla_\pi,~G_{\lambda\sigma}$ describes covariant derivative
and Einstein tensor, respectively, while
$\Box=g^{\lambda\sigma}\nabla_\lambda\nabla_\sigma$ corresponds
to d'Alembert's operator. Further, the quantities $f_R,~f_T$ and $f_Q$ represent the partial differentiation of
function $f$ with respect to $R,~T$ and $Q$, respectively. From  Eq.(\ref{3}), the expression of trace
is obtained as
\begin{align}\nonumber
&3\Box
f_R+\frac{1}{2}\Box(f_QT)-T(f_T+1)+\nabla_\pi\nabla_\rho(f_QT^{\pi\rho})+
R(f_R-\frac{T}{2}f_Q)\\\nonumber &+(Rf_Q+4f_T)\textit{L}_m
-2f+2R_{\pi\rho}T^{\pi\rho}f_Q -2
\frac{\partial^2\textit{L}_m}{\partial g^{\lambda\sigma}\partial
g^{\pi\rho}}\left(f_Tg^{\pi\rho}+f_QR^{\pi\rho}\right).
\end{align}
The case $Q=0$ boils down $f(R,T,Q)$ theory to $f(R,T)$
theory of gravity. However, the vacuum case of $f(R,T,Q)$ theory leads to $f(R)$ gravity
theory. It is important to note that $\textsl{L}_m$ is in the form
of second or superior orders, whereas in case of relativistic frame
which are connected by a specific matter collection, the second
variation of $\textsl{L}_m$ can be neglected. In frame work of
\cite{14}, the matter Lagrangian has no specific distinction for
ideal fluid, also second variation was taken to be negligible. In GR
prospective, Eq.(\ref{3}) can be demonstrated as
\begin{align}\label{4}
&R_{\lambda\sigma}-\frac{R}{2}
g_{\lambda\sigma}=G_{\lambda\sigma}={{T}_{\lambda\sigma}}^{\textrm{eff}},
\end{align}
where
\begin{align}\nonumber
{{T}_{\lambda\sigma}}^{\textrm{eff}}&=\frac{1}{(f_R-f_Q\textit{L}_m)}\left
[(f_T+\frac{1}{2}Rf_Q+1)T^{(m)}_{\lambda\sigma}+
\left\{\frac{R}{2}\left(\frac{f}{R}-f_R\right)-\textit{L}_mf_T-\frac{1}{2}\right.\right.\\\nonumber
&\left.\left.\times\nabla_{\pi}\nabla_{\rho}(f_QT^{\pi\rho})\right\}g_{\lambda\sigma}
-\frac{1}{2}\Box(f_QT_{\lambda\sigma})
-(g_{\lambda\sigma}\Box-\nabla_\lambda\nabla_\sigma)f_R-2f_QR_{\pi(\lambda}T^\pi_{~\sigma)}\right.\\\nonumber
&\left.+\nabla_\pi\nabla_{(\lambda}
[T^\pi_{~\sigma)}f_Q]+2\left(f_QR^{\pi\rho}+f_Tg^{\pi\rho}\right)\frac{\partial^2\textit{L}_m}{\partial
g^{\lambda\sigma}\partial g^{\pi\rho}}\right].
\end{align}

\subsection{Spherical Anisotropic Source}

We assume our relativistic stellar objects to be in spherical
shape with the following line element
\begin{equation}\label{5}
ds^2_-=-A^2(t,r)dt^{2}+B^2(t,r)dr^{2}+C^2(t,r)(d\theta^{2}
+\sin^2\theta{d\phi^2}),
\end{equation}
In spherical geometry, we consider the
distribution of anisotropic fluid which collapses adiabatically and
has mathematical form as
\begin{equation}\label{6}
T_{\lambda\sigma}=(P_{\bot}+\mu)V_{\lambda}V_{\sigma}+P_{\bot}g_{\lambda\sigma}
-\chi_\lambda\chi_\sigma(P_{\bot}-P_r),
\end{equation}
where, $P_{\bot}, ~P_r$ and $\mu$ describe the tangential, radial
pressure and energy density of the fluid, respectively. For comoving frame of reference, the
four-vectors are defined as
$V^{\lambda}=A^{-1}\delta^{\lambda}_{0}$ and
$\chi^{\lambda}=B^{-1}\delta^{\lambda}_{1}$, obeying
\begin{equation*}
V^{\lambda}V_{\lambda}=1,\quad\chi^{\lambda}\chi_{\lambda}=-1,\quad
\chi^{\lambda}V_{\lambda}=0.
\end{equation*}
The expansion scalar, ($\Theta=V^{\lambda}_{~~;\lambda}$), for our system leads to
\begin{equation}\label{7}
\Theta=\frac{1}{A}\left(\frac{\dot{B}}{B}
+\frac{2\dot{C}}{C}\right).
\end{equation}
The modified field equations for our line element associated
with matter distribution (\ref{6}) become
\begin{align}\nonumber
\mu^{\textrm{eff}}&=\frac{1}{(f_R+f_Q\mu)}\left[-\mu
f_T+\frac{f''_R}{B^2}+\left(\frac{2C'}{C}-\frac{B'}{B}\right)\frac{f'_R}{B^2}-\left(\frac{\dot{B}}{B}+\frac{2\dot{C}}{C}\right)
\frac{\dot{f_R}}{A^2}+\mu\chi_1\right.\\\nonumber
&\left.-\frac{R}{2}\left(\frac{f}{R}-f_R\right)+\chi_2\dot{\mu}+\chi_3\mu'+\frac{f_Q}{2A^2}\ddot{\mu}+\frac{f_Q}{2B^2}\mu''+\chi_4P_r
+\left(\frac{f'_Q}{B^2}-\frac{5}{2}f_QB'\right)\right.\\\label{8}
&\left.\times
P'_r+\frac{f_Q}{2B^2}P''_r-\frac{f_Q}{2A^2B}\dot{B}\dot{P_r}+\chi_5P_\bot-\frac{3f_Q}{2A^2C}\dot{P_\bot}\dot{C}
-\frac{3f_Q}{2B^2C}{P'_\bot}{C'}\right],\\\nonumber
P_r^{\textrm{eff}}&=\frac{1}{(f_R+f_Q\mu)}\left[\mu
f_T+\frac{\ddot{f_R}}{A^2}+\left(\frac{2\dot{C}}{C}-\frac{\dot{A}}{A}\right)\frac{\dot{f_R}}{A^2}-\left(\frac{A'}{A}
+\frac{2C'}{C}\right)\frac{f'_R}{B^2}+\chi_6P_r\right.\\\nonumber
&\left.+\frac{R}{2}\left(\frac{f}{R}-f_R\right)+\chi_7P'_r+\chi_8\dot{P_r}+\chi_9P_\bot+\frac{f_Q}{A^2}\frac{\dot{C}}{C}\dot{P_\bot}
-\frac{f_Q}{2A^2}\ddot{\mu}+\frac{f_Q}{2B^2}\frac{A'}{A}\mu'\right.\\\label{9}
&\left.+\chi_{10}\mu+\chi_{11}\dot{\mu}+\frac{f_Q}{B^2}\frac{C'}{C}P'_\bot\right],\\\nonumber
P_\bot^{\textrm{eff}}&=\frac{1}{(f_R+f_Q\mu)}\left[\mu
f_T+\frac{\ddot{f_R}}{A^2}-\frac{f''_R}{B^2}+\left(\frac{\dot{B}}{B}-\frac{\dot{A}}{A}+\frac{\dot{C}}{C}\right)\frac{\dot{f_R}}{A^2}
+\chi_{12}P_\bot+\chi_{13}\dot{P_\bot}\right.\\\nonumber
&\left.+\chi_{14}P'_\bot+\frac{R}{2}\left(\frac{f}{R}-f_R\right)+\left(\frac{B'}{B}-\frac{A'}{A}
-\frac{C'}{C}\right)\frac{f'_R}{B^2}-\frac{f_Q}{2A^2}\ddot{P_\bot}+\frac{f_Q}{2B^2}{P''_\bot}\right.\\\nonumber
&\left.+\chi_{16}\mu+\frac{f_Q}{2A^2}\frac{\dot{B}}{B}\dot{P_r}+\frac{5f_Q}{2B^2}\frac{B'}{B}P'_r-\frac{P'_rf'_Q}{B^2}-\frac{f_Q}{2B^2}P''_r
+\left(\frac{5f_Q\dot{A}}{2A^3}-\frac{\dot{f_Q}}{A^2}\right)\dot{\mu}\right.\\\label{10}
&\left.-\frac{f_Q}{2A^2}\ddot{\mu}+\frac{f_Q}{2B^2}\frac{A'}{A}\mu'+\chi_{15}P_r\right],\\\label{11}
T_{01}^{\textrm{eff}}&=\frac{1}{(f_R+f_Q\mu)}\left[\dot{f'_R}-\frac{A'}{A}\dot{f_R}-\frac{\dot{B}}{B}f'_R\right]\equiv
H,
\end{align}
where the quantities $\chi_i$'s consists of metric variables and
their derivatives, and their expressions are given in Appendix,
The derivatives with respect to time and radial coordinate are shown with the help of
$\cdot$ and $\prime$ notations, respectively. The value of the Ricci scalar $R$ is
given as
\begin{align}\nonumber
R=R(t,r)&=\left(\frac{2B'}{B}-\frac{C'}{C}-\frac{2A'}{A}\right)\frac{2C'}{CB^2}
-\frac{2}{B^2}\left(2\frac{C''}{C}-\frac{B'A'}{BA}+\frac{2A'}{A}\right)-2\frac{\dot{C}}{CA^2}\\\label{12}
&\times\left(\frac{2\dot{A}}{A}-\frac{\dot{C}}{C}-\frac{2\dot{B}}{B}\right)-\frac{2}{C^2}+\frac{2}{A^2}
\left(\frac{\ddot{B}}{B}-\frac{\dot{A}\dot{B}}{AB}+2\frac{\ddot{C}}{C}\right).
\end{align}

\section{The Perturbative Scheme and Collapse Equation}

Here, we wish to calculate $f(R,T,Q)$ field as well as dynamical
equations through perturbation technique. This technique would be
helpful to compute the instability zones for analytical models of
spherical geometry in $f(R,T,Q)$ theory of gravity.

\subsection{Mass Function and Divergence of Effective Energy Momentum Tensor}

Here, our aim is to explore the expression of hydrodynamical equation. For this purpose,
we shall define modified versions of dynamical equations and a viable
collapsing model in the framework. The mass function for spherical
geometry is found by using Misner-Sharp formalism as \cite{15}
\begin{align}\label{13}
m(t,r)=\frac{C}{2}\left(1-\frac{C'^2}{B^2}+\frac{\dot{C}^2}{A^2}\right).
\end{align}
Its radial derivative is obtained as
\begin{align}\nonumber
m'=\frac{C'C^2}{2A^2}\left(T_{00}^{\textrm{eff}}+\kappa\mu
A^2\right)-\frac{C^2\dot{C}}{2A^2}T_{01}^{\textrm{eff}}.
\end{align}
The integration of the above equation gives rise to
\begin{align}\nonumber
m=\frac{1}{2}\int^r_0\left(T_{00}^{\textrm{eff}}\frac{C'}{A^2}+\kappa\mu
C'-T_{01}^{\textrm{eff}}\frac{\dot{C}}{A^2}\right)C^2dr,
\end{align}
where we have considered  the case under which $m(t,0)=0$.

The divergence of effective energy momentum
tensor in this modified theory gives rise to
\begin{align}\label{14}
\nabla^\lambda
T_{\lambda\sigma}&=\frac{2}{Rf_Q+2f_T+1}\left[\nabla_\sigma(\textit{L}_mf_T)
+\nabla_\sigma(f_QR^{\pi\lambda}T_{\pi\sigma})-\frac{1}{2}(f_Tg_{\pi\rho}+f_QR_{\pi\rho})\right.\\\nonumber
&\times\left.\nabla_\sigma
T^{\pi\rho}-G_{\lambda\sigma}\nabla^\lambda(f_Q\textit{L}_m)\right],
\end{align}
This expression would gives us set of two equations as our matter
variables depend upon $t$ and $r$ variables. By using
$G^{\lambda\sigma}_{~;\sigma}=0$ and Eqs.(\ref{8})-(\ref{11}), this
correspondence with $\lambda=0,~1$ assign
\begin{align}\label{15}
&\left[\dot{\mu}^{\textrm{eff}}+\left(P_r^{\textrm{eff}}
+\mu^{\textrm{eff}}\right)\frac{\dot{B}}{B}+2(\mu^{\textrm{eff}}+P_\bot^{\textrm{eff}})\frac{\dot{C}}{C}\right]\frac{1}{A}
+AH'+AH\left(\frac{3A'}{A}+\frac{B'}{B}+\frac{2C'}{C}\right)=0,\\\label{16}
&\left[P'^{\textrm{eff}}_r+\left(P_r^{\textrm{eff}}
+\mu^{\textrm{eff}}\right)\frac{A'}{A}+2\left(P^{\textrm{eff}}_r
-P_\bot^{\textrm{eff}}\right)\frac{{C'}}{C}\right]\frac{1}{B}+B\dot{H}+BH
\left(\frac{\dot{A}}{A}+\frac{3\dot{B}}{B}+\frac{2\dot{C}}{C}
\right)=0.
\end{align}
The above dynamical equations could help to understand the hydrodynamics of
locally anisotropic spherical relativistic massive bodies. Here, the
superscript $\textrm{eff}$ shows the presence of $f(R,T,Q)$ dark
sources in the corresponding matter quantities. Now, we perform our analysis of
dynamical instability by using the following choice of $f(R,T,Q)$ formulations \cite{11} as
\begin{equation}\label{17}
f(R,T,Q)=\beta R(1+{\alpha}Q),
\end{equation}
in which the quantities $\alpha$ and $\beta$ are constant numbers. The particular values of these constants provides
modified correction for some particular cases. For instance, non-zero values of $\alpha$ and $\beta$
give rise to redefinition of gravitational field, thereby presenting this to be
physically viable model. First term in the model leads to
GR results.

\subsection{Perturbations}

The perturbation approach assists one to convert non-linear and
non-solvable relations to linear and solvable. This scheme is based
on the non-zero and very small perturbation parameter denoted by
$\epsilon$ with the assumption that $0<\epsilon\ll1$. We shall
perturb our equations up to first order in $\epsilon$. It is assumed that initially, the celestial system was
in the phase of hydrostatic equilibrium but with the passage of time it undergoes a periodic motion with frequency rate $\xi$. Therefore,
all the material and metric functions depend upon the time parameter
$\omega(t)$ at that instant. The perturbed configuration is
expressed as \cite{16}.
\begin{align}\label{18}
Y(t,r)=y_o(r)+{\epsilon}\omega(t)y(r),\quad
Z(t,r)=z_o(r)+\epsilon{\bar{z}}(t,r),
\end{align}
where $Y$ and $Z$ indicate the metric and material functions,
respectively. Applying this technique on Eq.(\ref{12}), the solution
of the second order partial differential equation can be written as
\begin{align}\label{19}
\omega=\omega(t)=-\exp({\xi} t),
\end{align}
where the frequency $\xi$ of the anisotropic spherical body using
Schwarzschild radius, is calculated as
\begin{align}\nonumber
\xi^2&=\frac{1}{r(2cB_o^2+br)}\left[\frac{2c'}{B_o^2}+\frac{2rbA'_o}{A_oB_o^3}+\frac{3a}{A_oB_o^2}
+\frac{2cA'_o}{A_oB_o^2}+\frac{2c'rA_o'}{A_oB_o^2}+\frac{2a'r}{A_oB_o^2}
-\frac{3b}{B_o}\right.\\\nonumber
&+\left.4a\frac{rA'_o}{A_o^2B_o^2}+\frac{b}{B_o^3}-\frac{2rb'}{B_o^3}
-\frac{2cB'_o}{B_o^3}-\frac{2rc'B'_o}{B_o^3}-\frac{6aB'_o}{A_oB_o^3}-2cr^2\frac{A_o'B_o'}{A_oB_o^3}+\frac{2rc''}{B_o^2}\right.\\\nonumber
&-\left.\frac{2ar^2A'_oB'_o}{A_o^2B_o^3}-\frac{b'r^2A'_o}{A_oB_o^3}-\frac{a'r^2B_o'}{A_oB_o^3}
+\frac{ar^2A''_o}{A_o^3B_o^2}+\frac{r^2a''}{A_o^2B_o^2}+\frac{br^2A_o''}{A_o^2B_o^3}
-\frac{3a}{A_o}+\frac{2crA_o''}{A_o^2B_o^2}\right.\\\label{20}
&+\left.\frac{R_o}{2}\left(\frac{3a}{B_o^3A_o^4}+\frac{3b}{B_o^4A_o^3}
+\frac{2c}{rA_o^3B_o^3}\right)\right]B_o^2A_o^2 .
\end{align}
By using Eqs.(\ref{18}), the perturbed $f(R,T,Q)$ gravity model, is
\begin{align}\label{21}
f=R_o(1+\alpha Q_o)+\epsilon \omega(t)[d+\alpha(R_og+dQ_o)],
\end{align}
where, $d=d(r),~g=g(r)$. The static background of $f(R,T,Q)$ field equations
obtained from perturbation scheme are
\begin{align}\nonumber
\mu^{\textrm{eff}}_o&=\frac{1}{1+\alpha(Q_o+\mu_oR_o)}\left[\frac{\alpha}{B_o^2}\left(Q''_o-\frac{B'_oQ'_o}{B_o}+\frac{2Q'_o}{r}\right)
+\mu_o\chi_{1o}+\mu'_o\chi_{3o}\right.\\\label{22} &+\left.P_{\bot
o}\chi_{5o}+\frac{\alpha
R_o}{2B_o^2}\left(\mu''_o+P''_{ro}-\frac{3P'_{\bot}}{r}
+2P'_{ro}\frac{R'_o}{R_o}\right)-\frac{5\alpha}{2}R_oB'_oP'_{ro}\right],\\\nonumber
P^{\textrm{eff}}_{ro}&=\frac{1}{1+\alpha(Q_o+\mu_oR_o)}\left[-\frac{\alpha
Q'_o}{B_o^2}\left(\frac{A'_O}{A_O}+\frac{2}{r}\right)
+\mu_o\chi_{10o}+P_{ro}\chi_{6o}+P'_{ro}\chi_{7o}\right.\\\label{23}
&+\left.P_{\bot
o}\chi_{9o}+\frac{\alpha}{2B_o^2}\left(\mu'_oR'_o\frac{A'_o}{A_o}+2R_o\frac{P'_{\bot
o}}{r}\right)\right],\\\nonumber P^{\textrm{eff}}_{\bot
o}&=\frac{1}{1+\alpha(Q_o+\mu_oR_o)}\left[-\frac{\alpha}{B_o^2}\left(Q''_o+Q'_o(\frac{A'_o}{A_o}-\frac{B'_o}{B_o}+\frac{1}{r})\right)
+\mu_o\chi_{16o}\right.\\\label{24} &\left.+P_{ro}\chi_{15o}+P_{\bot
o}\chi_{12o}+P'_{\bot o}\chi_{14o}+\frac{\alpha
R_o}{2B_o^2}(P''_{\bot o}-P''_{ro}
+5P'_{ro}\frac{B'_o}{B_o}\right.\\\nonumber
&+\left.\mu'_o\frac{A'_o}{A_o}-2P'_{ro}\frac{R'_o}{R_o})\right],
\end{align}
while the non-static field equations under this strategy have the following form
\begin{align}\nonumber
\bar{\mu}^{\textrm{eff}}&=\frac{1}{1+\alpha(Q_o+\mu_o
R_o)}\left[\frac{\alpha\omega}{B_o^2}\left\{g''-2b\frac{Q''_o}{B_o}-Q'_o(\frac{b}{B_o})'+2b\frac{Q'_oB'_o}{B_o^2}
-4b\frac{Q'_o}{rB_o}\right.\right.\\\nonumber
&\left.\left.+2Q'_o(\frac{c}{r})'-g'\frac{B'_o}{B_o}+2\frac{g'}{r}+d\frac{\mu''_o}{2}+P'_{ro}d'-\frac{3d}{2r}P'_{\bot
o}+d\frac{P''_{ro}}{2}\right\} +\omega(\mu_ox_1\right.\\\nonumber
&+\mu'_ox_3+P_{ro}x_4+P_{\bot
o}x_5)+\bar{\mu}\chi_{1o}+\dot{\bar{\mu}}\chi_{2o}+\bar{\mu}'\chi_{3o}+\bar{P_r}\chi_{4o}+\bar{P_\bot}\chi_{5o}\\\nonumber
&\left. +\frac{\alpha
R_o}{B_o^2}\left\{\frac{\bar{\ddot{\mu}}}{A_o^2}+\frac{\bar{\mu}''}{B_o^2}+\frac{\bar{P_r}''}{2}\right\}
-\frac{\alpha\omega
bR_o}{B_o^2}\left\{\frac{\mu''_o}{B_o}+2\frac{P'_{ro}R'_o}{R_o}+\frac{P''_{ro}}{B_o}
-3\frac{{P_{\bot o}'}}{rB_o}\right.\right.\\\nonumber
&\left.+\frac{3P_{\bot
o}'}{2b}\left(\frac{c}{r}\right)'\right\}\left.-\frac{\alpha\omega}{2}\left(5b\frac{dB'_oP'{ro}}{2}+5b'R'_oP'_{ro}\right)-\frac{\alpha
R_o}{2}\left\{5B'_o\bar{P'_r}
+3\frac{\bar{P_\bot'}}{rB_o^2}\right.\right.\\\label{23}
&\left.\left.+2\frac{R'_o\bar{P'_r}}{B_o^2R_o}\right\}
-\alpha\mu^{\textrm{eff}}_o\left(R_o\bar{\mu}+\omega(g+d\mu_o)\right)\right],\\\nonumber
\bar{P_r}^{\textrm{eff}}&=\frac{1}{1+\alpha(Q_o+\mu_oR_o)}\left[\frac{\alpha\ddot{\omega}}{B_o^2}+\frac{\alpha\omega}{B_o^2}\left(
2b\frac{Q'_oA'_o}{B_oA_o}-Q'_o(\frac{a}{A_o})'+4b\frac{Q'_o}{rB_o}-2Q'_o(\frac{c}{r})'\right.\right.\\\nonumber
&\left.\left.-\frac{g'A'_o}{A_o}-\frac{2g'}{r}+d\frac{\mu'_oA'_o}{2A_o}+d\frac{P_{\bot
o}'}{r}\right) +\omega\left(P_{ro}x_6+P'{ro}x_7+P_{\bot
o}x_9+\mu_ox_{10}\right)\right.\\\nonumber
&\left.+\bar{P_r}\chi_{6o}+\bar{P'_r}\chi_{7o}
+\dot{\bar{P_r}}\chi_{8o}+\bar{P_\bot}\chi_{9o}+\bar{\mu}\chi_{10o}+\dot{\bar{\mu}}\chi_{11o}+\frac{\alpha\omega
R_o}{B_o^2}
\left\{\frac{\mu'_o}{2}(\frac{a}{A_o})'\right.\right.\\\nonumber
&-\left.\left.b\mu'_o\frac{A'_0}{B_oA_o}+P_{\bot
o}'(\frac{c}{r})'-2b\frac{P_{\bot o}'}{rB_o}\right\} -\frac{\alpha
R_o}{2}\left\{\frac{\ddot{\bar{\mu}}}{A_o^2}-\frac{\bar{\mu}'A'_o}{B_o^2A_o}-2\frac{\bar{P_{\bot
o}'}}{rB_o^2}\right\} -\alpha
P^{\textrm{eff}}_{ro}\right.\\\nonumber
&\times\left.\left(R_o\bar{\mu}+\omega(g+d\mu_o)\right)\right],\\\nonumber
\bar{P_\bot}^{\textrm{eff}}&=\frac{1}{1+\alpha(Q_o+\mu_oR_o)}\left[\frac{\alpha\ddot{\omega}}{A_o^2}
-\frac{\alpha\omega}{B_o^2}\left\{g''-2b\frac{Q''_o}{B_o}-2b\frac{Q'_oA'_o}{B_oA_o}+Q'_o(\frac{a}{A_o})'\right.\right.\\\nonumber
&+\left.\left.3b\frac{Q'_oB'_o}{B_o^2}-\frac{b'Q'_o}{B_o}-2\frac{b
Q'_o}{rB_o}+Q'_o(\frac{c}{r})'+g'(\frac{A'_o}{A_o}-\frac{B'_o}{B_o}+\frac{1}{r})\right\}+\omega\left(P_{\bot
o}x_{12}\right.\right.\\\nonumber &\left.\left.+P_{\bot
o}'x_{14}+P_{ro}x_{15}+\mu_ox_{16}\right) +\bar{P_{\bot
o}}\chi_{12o}+\dot{\bar{P_\bot o}}\chi_{13o}+\bar{P_{\bot
o}'}\chi_{14o}+\bar{P_r}\chi_{15o}\right.\\\nonumber
&\left.+\bar{\mu}\chi_{16o} -\frac{\alpha
R_o}{2}\left\{\frac{\ddot{\bar{\mu}}}{A_o^2}+\frac{\ddot{\bar{P_\bot
}}}{A_o^2}-\frac{\bar{P_{\bot o}''}}{B_o^2}+
\frac{\bar{P_r}''}{B_o^2}-\frac{\bar{\mu}'A'_o}{B_o^2A_o}-5\frac{\bar{P_r}'B'_o}{B_o^3}\right\}
-\frac{\alpha\omega R_o}{B_o^2}\right.\\\nonumber
&\times\left.\left\{\frac{bP_{\bot
o}''}{B_o}-2b\frac{P_{ro}'R'_o}{B_oR_o}
-\frac{bP_{ro}''}{B_o}+b\mu'_o\frac{A'_o}{B_oA_o}-\frac{\mu'_o}{2}(\frac{a}{A_o})'-\frac{5P_{ro}'}{2B_o}(b'+b
B'_o)\right\}\right.\\\nonumber
&\left.+\frac{\alpha\omega}{B_o^2}\left\{\frac{d}{2}(P_{\bot
o}''-P_{ro}''+\frac{\mu'_oA'_o}{A_o}+5\frac{B'_oP_{ro}'}{B_o})-d'P_{ro}'\right\}\right.\\\nonumber
&\left.-\alpha P^{\textrm{eff}}_{\bot
o}\left(R_o\bar{\mu}+\omega(g+d\mu_o)\right)\right].
\end{align}
In hydrostatic equilibrium position, the second conservation law has the
static form
\begin{align}\nonumber
\frac{1}{B_o}\left[P'^{\textrm{eff}}_{ro}+\left(\mu^{\textrm{eff}}_o+P^{\textrm{eff}}_{ro}\right)\frac{A'_o}{A_o}
-\frac{2}{r}\left(P^{\textrm{eff}}_{\bot
o}-P^{\textrm{eff}}_{ro}\right)\right]=0.
\end{align}
The non-static perturbed configurations of Eqs.(\ref{15}) and
(\ref{16}) are
\begin{align}\nonumber
&\frac{1}{A_o}\left[\dot{\bar{\mu}}^{\textrm{eff}}+\frac{b\dot{\omega}}{B_o}
\left(\mu^{\textrm{eff}}_o+P^{\textrm{eff}}_{ro}\right)
+2\frac{c\dot{\omega}}{r}\left(P^{\textrm{eff}}_{\bot
o}+\mu^{\textrm{eff}}_{o}\right)\right]+\omega A_oh'\\\label{25} &+
h\omega\left(\frac{2A_o}{r}+3A_o'\right)=0,\\\nonumber
&\frac{1}{B_o^2}\left[\bar{P}'^{\textrm{eff}}_{r}+\frac{A_o'}{A_o}(\bar{P}^{\textrm{eff}}_{r}+
\bar{\mu}^{\textrm{eff}})+\left(\frac{a}{A_o}\right)'({P}^{\textrm{eff}}_{ro}+
{\mu}^{\textrm{eff}}_o)\omega-2\omega\left(\frac{c}{r}\right)' (
{P}^{\textrm{eff}}_{\bot
o}-{P}^{\textrm{eff}}_{ro})\right.\\\label{26}
&-\left.\frac{2}{r}(\bar{P}^{\textrm{eff}}_{\bot}-\bar{P}^{\textrm{eff}}_{r})\right]
+h\dot{\omega}=0.
\end{align}
From Eq.(\ref{11}), we can find the relation between $b$ and $g$ as follows
\begin{align}\nonumber
\frac{\alpha\dot{\omega}}{1+\alpha(Q_o+\mu_o
R_o)}\left[g'-\frac{gA_o'}{A_o}-\frac{bQ'_o}{B_o}\right]=0.
\end{align}
In spherical geometry, the matter content described by Misner and
Sharp, in static and non-static positions give
\begin{align}\label{27}
m_o=m_o(r)&=\left(1-\frac{1}{B_o^2}\right)\frac{r}{2},\quad
\bar{m}=\bar{m}(t,r)=\frac{r\omega}{B_o^2}\left\{\left(\frac{b}{B_o}-c'\right)-\frac{c}{2r}(1-B_o^2)\right\}.
\end{align}

\subsection{Stability Analysis}

Dynamical stability is the characteristics of an object or system to
retain its stable position, whenever it is subjected to
perturbations. The dynamical stability has utmost relevance in
structure formation and evolution of self-gravitating bodies. The
instability/stability of celestial bodies has been discussed not
only in the framework of general relativity but also in different
modified gravity theories. In order to analyze the dynamics of
massive objects, one can calculate its stability condition with N
and pN approximations. It is interesting to analyze what happens
when the phase of equilibrium of stellar structures is disturbed?
Will this perturbation be relaxed (stable state) or will it grow
(unstable state). In this respect, one needs to take into account
the dynamical instability problem or thermal instability issue.

It is found that under hydrostatic equilibrium phase, the stability
criterion can easily be achieved by making linearized field
equations as well as conservation equations against radial
perturbation. We remark that the realistic object moves, during
evolution, via several evolutionary phases determined by
instability/stability degrees of freedom. This suggests that
relativistic systems can be stable at one instant but not at the
other. Thus, one needs to understand the dynamical behavior of
self-gravitating systems by calculating instability regions at both
N as well as pN regimes. Such epochs have vital role in the
discussion of gravitational collapse of compact objects.

Now, we discuss the stability of local anisotropic spherical dense
objects using the equations developed in the previous section. One can
understand the notion of instability of relativistic interiors via
adiabatic index ($\Gamma_1$). Equation of state suggested by
Harrison et al. \cite{16} provides a relationship between energy density
and pressure of the source which measures a change in pressure
corresponding to a given change in energy density.
This is
\begin{equation}\label{28}
\bar{P_i}=\Gamma_1\frac{P_{i0}}{\mu_0+P_{i0}}\bar{\mu}.
\end{equation}
We have adopted Harrison et al. equation of state \cite{16} as it measures the stiffness of the fluid.
As we are interested in examining the role played by matter variables on the stability of spherical system
in the background of a particular modified gravity. Therefore, we have chosen such equation of state with the assumption that
adiabatic index is constant throughout the matter distribution or,
at least, within the observed region of spacetime. Also, we are exploring the stability conditions for non-static anisotropic spherical geometry.
For this purpose, we take the equation of state in the
scenario of second law of thermodynamics, where both $\bar{P}$ and $\bar{\mu}$
are functions of $t$ and $r$ obtained after first order perturbation.
It is not possible to use this scheme to investigate the instability range
with an equation of state which is not perturbed or does not contain non-static terms.
Consequently, the other equation of state would lead to
different kind of investigation for stability conditions which have also been discussed in literature \cite{eos1}.
Equation (\ref{28}) can be re-casted as
\begin{align}\label{29}
\dot{\bar{\mu}}^{\textrm{eff}}=-\left[\frac{b}{B_o}\left(P^{\textrm{eff}}_{ro}+\mu^{\textrm{eff}}_o\right)
+\frac{2c}{r}\left(\mu^{\textrm{eff}}_{o}+P^{\textrm{eff}}_{\bot
o}\right)\right]\dot{\omega}-\omega J_1,
\end{align}
where
\begin{align}\nonumber
J_1=h'-3h\frac{A'_o}{A_o}-\frac{2h}{r}.
\end{align}
Using value of $B_o^2=\frac{r}{r-2m_o}$, in hydrostatic part of 11
field equation, we have
\begin{align}\nonumber
&\frac{A_o'}{A_o}=\frac{1}{(r-2m_o)
(\frac{\alpha}{2}\mu'_oR'_o-\alpha
Q'_o)}\left[P^{\textrm{eff}}_{ro}(r+\alpha r(Q_o+\mu_oR_o))+2\alpha
Q'_o\right.\\\nonumber &-4\alpha Q'_o\frac{m_o}{r}-r (
p_{ro}\chi_{6o}\left.+ p'_{ro}\chi_{7o}+ p_{\bot
o}\chi_{9o}+\mu_o\chi_{10o}) -\alpha R_o p'_{\bot o}+2\alpha
R_op'_{\bot o}\frac{m_o}{r}\right].
\end{align}
The static profile of the Ricci scalar is
\begin{align}\nonumber
R_o&=\left(1-\frac{1}{B_o^2}\right)\frac{2}{r^2}-\frac{2A_o'}{A_oB_o^2}
\left(\frac{2}{r}-\frac{B_o'}{B_o}\right)^2-\frac{2}{B_o^2}\left(\frac{A_o''}{A_o^2}-\frac{2B_o'}{rB_o}\right).
\end{align}
Taking integration of Eq.(\ref{29}) with $t$, we have
\begin{align}\label{30}
\bar{\mu}^{\textrm{eff}}=-J\omega,
\end{align}
where
\begin{align}\nonumber
J=\left[\frac{b}{B_o}\left(P^{\textrm{eff}}_{ro}+\mu^{\textrm{eff}}_o\right)
+\frac{2c}{r}\left(\mu^{\textrm{eff}}_{o}+P^{\textrm{eff}}_{\bot
o}\right)\right]+\frac{J_1}{\xi}.
\end{align}
Putting the expression of $\bar{\mu}^{\textrm{eff}}=-J\omega$ in
Eq.(\ref{28}), one can find
\begin{align}\label{31}
\bar{P}^{\textrm{eff}}_{r}=-\Gamma_1
\frac{P^{\textrm{eff}}_{ro}J\omega}{(\mu^{\textrm{eff}}_o+P^{\textrm{eff}}_{ro})},\quad
\bar{P}^{\textrm{eff}}_{\bot}=-\Gamma_1 \frac{P^{\textrm{eff}}_{\bot
o}J\omega}{(\mu^{\textrm{eff}}_o+P^{\textrm{eff}}_{\bot o})}.
\end{align}
By making use of Eqs.(\ref{30}), (\ref{31}) and (\ref{26}), the corresponding hydrodynamical equation turns out to be
\begin{align}\nonumber
&\frac{2\omega}{B_o}\left(P^{\textrm{eff}}_{ro}-P^{\textrm{eff}}_{\bot
o}\right)\left(\frac{c}{r}\right)'-\Gamma_1\frac{J\omega}{B_o}
\frac{P'^{\textrm{eff}}_{ro}}{(\mu^{\textrm{eff}}_o+P^{\textrm{eff}}_{r
o})}+\Gamma_1 \frac{\omega J}{B_o}
\frac{\mu'^{\textrm{eff}}_oP^{\textrm{eff}}_{ro}}{(\mu^{\textrm{eff}}_o+P^{\textrm{eff}}_{ro})^2}
+\Gamma_1\frac{J\omega}{B_o}\\\nonumber
&\times\frac{P^{\textrm{eff}}_{ro}P'^{\textrm{eff}}_{ro}}{(\mu^{\textrm{eff}}_o
+P^{\textrm{eff}}_{ro})^2}-\frac{A'_oJ\omega}{A_oB_o}-
\frac{A'_oJ\omega}{A_oB_o}\frac{P^{\textrm{eff}}_{ro}}
{(\mu^{\textrm{eff}}_o+P^{\textrm{eff}}_{ro})}\Gamma_1+\frac{\omega}{B_o}
\left(\mu^{\textrm{eff}}_o+P^{\textrm{eff}}_{ro}\right)\left(\frac{a}{A_o}\right)'\\\label{32}
&-\Gamma_1\frac{J'\omega}{B_o}
\frac{P^{\textrm{eff}}_{ro}}{(\mu^{\textrm{eff}}_o+P^{\textrm{eff}}_{ro})}+
\frac{2\omega
J}{rB_o}\Gamma_1\frac{P^{\textrm{eff}}_{ro}}{(\mu^{\textrm{eff}}_o+
P^{\textrm{eff}}_{ro})}-\frac{2\omega
J}{rB_o}\frac{P^{\textrm{eff}}_{\bot o}}{(\mu^{\textrm{eff}}_o+
P^{\textrm{eff}}_{\bot o})}\\\nonumber
&+(\xi\omega)^2 \left(\alpha
g'-\alpha(g\frac{A'_o}{A_o}+ b\frac{Q'_o}{B_o}) \right)=0.
\end{align}
The above equation is also known as modified version of collapse
equation, in which the matter variables are related with stiffness parameter
This equation yield the effects of counter gravity and pressure gradients in a single expression.
The rest of the entries are the originator of the gravity forces.
The effects, generated by $f(R,T,Q)$ gravity terms and principal
stresses mediated by perfect fluid are of having pivotal role in order to understand
gravitational forces. We will analyze the collapse rate of the
celestial model in spherical geometry, in that case the dynamical
quantity, i.e., $\Gamma_1$ is positive only which would make the
stable hydrostatic environment among gravitational forces and principal stresses.

\section{Newtonian Approximations}

Now, we find the stability conditions for the spherical locally
anisotropic interiors with N limit. For this purpose, we take flat
background metric which leads to weak field approximations as
follows
\begin{equation*}
\mu_0\gg P_{i0}, \quad A_0=1, \quad B_0=1.
\end{equation*}
We have assumed that the contribution of energy density of matter
distribution is much much greater than its pressure components. In
order to achieve the stability regions of anisotropic spherical
compact stellar system, we need to consider that each term in the
the collapse equation to be positive then the expression of
corresponding hydro-dynamical equation takes the form
\begin{align}\label{33}
\Gamma_1F=a'\mu^{\textrm{eff}}_{o} -\xi_N^2(\alpha g'-\alpha b Q_o),
\end{align}
where $Z_N=\frac{J_{1N}}{\xi_N}$ with
\begin{align}\nonumber
F&=\left[\frac{2}{r}\left(b+\frac{2c}{r}\right)\left(P^{\textrm{eff}}_{ro}-P^{\textrm{eff}}_{\bot
o}\right)-\left(b+\frac{2c}{r}\right)P'^{\textrm{eff}}_{ro}-P^{\textrm{eff}}_{ro}Z_N\right],\\\nonumber
\xi_N&=\left[\frac{4}{r^3(br+2c)}\left(c'+a'r-rb'+c''r
+\frac{a''r^2}{2}-b\right)\right]^{\frac{1}{2}},
\end{align}
and $J_1$ consists the dark sources terms mediated from $\beta
R(1+{\alpha}Q)$ model. Equations (\ref{33}) provides the following
values of adiabatic index
\begin{align}\label{34}
\Gamma_1=\frac{a'\mu^{\textrm{eff}}_o-\alpha\omega\xi_N(g'-bQ'_o)}{F}.
\end{align}
This provides the hydrostatic condition which implies that the
system enters in the stable window for
\begin{equation}\label{35}
\Gamma_1>\frac{a'\mu^{\textrm{eff}}_o-\alpha\omega\xi_N(g'-bQ'_o)}{F}.
\end{equation}
In order to keep $\Gamma_1>0$, we need to consider
$|a'\mu^{\textrm{eff}}_o-\alpha\omega\xi_N(g'-bQ'_o)|$ and $|F|$.
Thus the system remains in the stable phase as long as it obeys
inequality (\ref{35}). This represents that the instability range
(\ref{35}), that has been calculated through adiabatic index,
depends upon the pressure components as well as anti-gravitational
force coupled with adiabatic index and gravitational force. These
variable quantities eventually depends upon the radial profiles of
the energy density, anisotropicity and $f(R,T,Q)$ curvature terms.
It is pertinent to note that the presence of anisotropicity in the
fluid pressure has greatly influenced the instability regimes of the
spherical relativistic structure at N epoch as described by
inequality (\ref{35}). By keeping the absolute values of
denominator, we noticed that effective pressure anisotropicity tend
to decreases the stability regions or tends to remove hindrances for
the system to move in the collapsing phase. This result is
well-consistent with \cite{ck2}. One can easily notice that
$f(R,T,Q)$ terms appearing in the expression (\ref{35}) tends to
decrease the stability range as these some of these terms are
appearing in the numerator with negative sign. It has been seen that
this expression contains effective forms of matter variables that
shows that terms coming from the coupling of matter and geometry
have greatly modify the instability constraints due to their
non-attractive nature. We now briefly describe our results as
follows:
\begin{enumerate}
\item If the gravitational forces $|a'\mu^{\textrm{eff}}_o-\alpha\omega\xi_N(g'-bQ'_o)|$ are balanced by the anti-gravitational
and effective pressure forces $|F|$, (thereby boiling down
inequality (\ref{34}) to $\Gamma_1=1$) then the system will rest in
the window of hydrostatic equilibrium.
\item If the modified gravity forces produced by
$|a'\mu^{\textrm{eff}}_o-\alpha\omega\xi_N(g'-bQ'_o)|$ are greater
than that of $|F|$, then the system will enter in the stable phase
instead of collapsing, i.e, counter gravitational forces as well as
effective principal pressures give the stability constraint
$\Gamma_1>1$.
\item The celestial system will be in unstable state whenever
it achieve contribution from
$|\mu'^{\textrm{eff}}_o-\alpha\omega\xi_N(g'-bQ'_o)|$ to be lesser
than from $|F|$. This assigns the range of $\Gamma_1$ belonging to
the open interval $(0,1)$.
\end{enumerate}
The GR limit, $f(R,T,Q)=R$, converts all the effective fluid
variables appearing in inequality (\ref{35}) to usual matter
variables, i.e., $P^{\textrm{eff}}_{ro}\rightarrow
P_{ro},~P^{\textrm{eff}}_{\bot o} \rightarrow P_{\bot o}$ and
$\mu^{\textrm{eff}}_{o}\rightarrow \mu_{o}$. Furthermore, under this
limit, the quantity $Z_N$ vanishes, thereby recovery the whole
dynamics in the framework of GR. Thus, the stability constraint
becomes
\begin{align}\label{35a}
\Gamma_1>\frac{|a'\mu_o|}{\frac{2}{r}(P_{ro}-P_{\bot
o})(b+\frac{2c}{r})-\left(b+\frac{2c}{r}\right)P'_{ro}}.
\end{align}
This expression exactly match with that obtained in \cite{zzz2},
under certain conditions. However, for isotropic spherical system, the
stability constraint with N approximations boils down to
\begin{align}\label{35a}
\Gamma_1>\frac{|a'\mu_o|}{|\left(b+\frac{2c}{r}\right)P'_{o}|}.
\end{align}

\section{Post-Newtonian Approximation}

Different aspects of various gravitational framework can upraise
particular problems in applications of practical interest. These
contain the nonlinear nature of equations of motion and the
non-existence of a background geometry that can be utilized to
discuss physically interesting quantities, for example, energy and
momentum. In this direction, some approximations techniques are
applied to construct physically concerning predictions. The example
of such approximation scheme is a linearized gravity where the non
linear parts of spacetime metric are ignored, that eventually give
rise to some useful approximate outcomes. As a result of this
scheme, linearized field equations describing weak gravitational
field can easily be governed. Thus, N and pN limits are considered
as the approximations for the weak field of relativistic
gravitational theory, in which the corresponding equations of motion
and metric variables are approximated in the inverse power of the
light speed.

In the realm of gravitational theories, both N and pN approximations
narrate the order of small perturbations/deviations of any local
system from its isotropic flat and homogeneous environment. One can
evaluate these approximations by expanding metric functions through
Taylor series as \cite{zzz3}
$$g_{\gamma\delta}\approx\eta_{\gamma\delta}+\hbar_{\gamma\delta}~, \quad |\hbar_{\gamma\delta}|\ll1$$
with
$$\hbar_{00}\approx\hbar^{(II)}_{00}+\hbar^{(IV)}_{00}+...~,\quad \hbar_{0i}\approx\hbar^{(III)}_{0i}+\hbar^{(V)}_{0i}+...~,\quad \hbar_{ij}\approx\hbar^{(II)}_{ij}+\hbar^{(IV)}_{ij}+..., ~~i,j=
1,2,3$$ where the superscripts $(II),~(III),~(IV)$ describe approximation
orders up to
$\left(\frac{1}{c^2}\right),~\left(\frac{1}{c^3}\right)$ and
$\left(\frac{1}{c^4}\right)$, while $\eta_{\mu\nu}$ stands for the
Minkowski spacetime that represents isotropic and homogeneous flat
environment of $g_{\gamma\delta}$ and $\hbar_{\mu\nu}$ indicates
perturbation of metric tensor $g_{\mu\nu}$ from $\eta_{\mu\nu}$
(background values). The approximations
$g_{00}\sim\eta_{00}+\hbar^{(II)}_{00}~,g_{ij}\sim\eta_{ij}$ yields N
limit, while the pN limits need the information of
$g_{00}\sim\eta_{00}+\hbar^{(II)}_{00}+\hbar^{(IV)}_{00}~,g_{0i}\sim\hbar^{(III)}_{0i}$
along with $g_{ij}\sim\eta_{ij}+\hbar^{(II)}_{ij}$. This describes
the peculiar connection between the calculations of N and pN limits
for any toy relativistic model.

To achieve pN instability limits, we take
$A_0(r)=1-\phi,~B_0(r)=1+\phi$, with linear $O(\phi)$ and
$\phi(r)=\frac{m_0}{r}$. In this aspect, the value of stiffness
parameter through the collapse equation can be given as
\begin{align}\label{pn1}
\Gamma_1&=\frac{E_{pN}}{\psi J'_{pN}-kJ_{pN}},
\end{align}
where
\begin{align}\nonumber
E_{pN}&=P^{\textrm{eff}}_{ro}\left(\gamma-2 (\frac
{c}{r})'(1-\frac{2m_o}{r})\right) +(\mu^{\textrm{eff}}_o
\left(\gamma-\frac{2c}{r}(\frac{m_o}{r})'(1-\frac{m_o}{r})\right)+P^{\textrm{eff}}_{\bot
o}\left[\right.((\frac{m_o}{r})'(-\frac{2m_o}{r}))Z_{pN}\\\nonumber
&-\frac{2c}{r}(\frac{m_o}{r})'(1-\frac{m_o}{r})
-2(\frac{c}{r})'(1-\frac{2m_o}{r})-\alpha\omega\xi^2_{pN}\left(g'+g(\frac{m_o}{r})'
(1+\frac{m_o}{r})-bQ'_o(1-\frac{m_o}{r})\right)\left.\right]
\end{align}
The locally anisotropic spherical compact structure will move in the
window of stable regime, if modified gravitational forces mediated
by $|\psi J'_{pN}-kJ_{pN}|$ are lesser than produced by $|E_{pN}|$.
Then, the stability of the spherical fluids at pN epoch can be
checked through
\begin{align}\label{pn2}
\Gamma_1>\frac{E_{pN}}{\psi J'_{pN}-kJ_{pN}}.
\end{align}
If the relativistic interior is able to accomplish the state
satisfying $|E|=|\psi J'_{pN}-kJ_{pN}|$, then the system reverts
itself in its initial hydrostatic equilibrium state. This situation
can be dealt with the help of Eq.(\ref{pn1}). The system will enter
into dynamical instability window, if the influence of $|E|$ is less
than $|\psi J'_{pN}-kJ_{pN}|$. This gives rise to
\begin{align}\nonumber
\Gamma_1<\frac{E_{pN}}{\psi J'_{pN}-kJ_{pN}}.
\end{align}
For the case of isotropic spherically symmetric system, the instability regime
with pN approximations turns out to be
\begin{align}\nonumber
\Gamma_1<\frac{\mathbb{E}_{pN}}{\bar{\psi} \mathbb{J}'_{pN}-k\mathbb{J}_{pN}},
\end{align}
where
\begin{align}\nonumber
\mathbb{J}_{pN}&=\left[\left\{b\left(1+\frac{2m_o}{r}\right)+\frac{2c}{r}\right\}\left(P^{\textrm{eff}}_{o}+\mu^{\textrm{eff}}_o\right)\right]
+\frac{J_1}{\xi},\quad \bar{\psi}=\frac{r-2m_o}{r(\mu_o+P_{o})}P_{o}\\\nonumber
\mathbb{E}_{pN}&=\mu^{\textrm{eff}}_o
\left[\gamma-\frac{2c}{r}\left(\frac{m_o}{r}\right)'\left(1-\frac{m_o}{r}\right)\right]+P^{\textrm{eff}}_{
o}\left[\left(\frac{m_o}{r}\right)'\left(-\frac{2m_o}{r}\right)Z_{pN}-4\left(\frac{c}{r}\right)'\left(1-\frac{2m_o}{r}\right)\right.\\\nonumber
&\left.-\frac{2c}{r}\left(\frac{m_o}{r}\right)'\left(1-\frac{m_o}{r}\right)
+\gamma-\alpha\omega\xi^2_{pN}\left\{g'+g\left(\frac{m_o}{r}\right)'
\left(1+\frac{m_o}{r}\right)-bQ'_o\left(1-\frac{m_o}{r}\right)\right\}\right].
\end{align}
Under GR limit, i.e., when $f(R,T,Q)=R$, the instability limit
provided by the collapse equation through $\Gamma_1$ at pN limits
boils down to
\begin{align}\label{36}
\Gamma_1&<\frac{E_{GR}}{\psi J'_{pN}-kJ_{pN}}
\end{align}
where
\begin{align}\nonumber
E_{GR}&=P_{ro}\left(\gamma-2 (\frac
{c}{r})'(1-\frac{2m_o}{r})\right) +\mu_o
\left\{\gamma-\frac{2c}{r}(\frac{m_o}{r})'\left(1-\frac{m_o}{r}\right)\right\}',\\\nonumber
&+P_{\bot o}\left[\right.
-\frac{2c}{r}(\frac{m_o}{r})(1-\frac{m_o}{r})
-2(\frac{c}{r})'(1-\frac{2m_o}{r})],\\\nonumber
J_{pN}&=\frac{b}{1+\frac{m_o}{r}}\left[P_{ro}+\mu_o+\frac{2c}{r}(\mu_o+P_{ro})\right],\\\nonumber
\gamma&=-b\left(\frac{m_o}{r}\right)'\left(1-\frac{2m_o}{r}\right)-a'\left(1+\frac{m_o}{r}\right)-a\left(\frac{m_o}{r}\right)',\\\nonumber
\psi&=\frac{r-2m_o}{r(\mu_o+P_{ro})}P_{ro},\quad Z_{pN}=0.
\end{align}
This reveals the significance of static profiles of relative matter
variables and stiffness parameter. This constraint exactly match
with that already obtained in \cite{zzz2} under certain conditions.

\section{Conclusions}

The stability problem of dense objects in the field of
modified gravitational has come into sight as a main concern. In this setting,
the instability eras for the locally anisotropic self-gravitating spherical configurations
are examined with a particular formulation $f(R,T,Q)$ gravity. We
have calculated the corresponding equations of motion for the locally anisotropic matter filled in spherical
irrotational symmetry. The conservation laws are
explored after using the contracted formulations of Bianchi identities with the background of effective energy momentum tensor.
The radial perturbation scheme is applied on main equations and then static as well as non static profiles of
field and dynamical expressions are presented.

We first assume our relativistic sphere rests in the window of hydrostatic
phase at the initial times. But, as time passes, the evolving system starts to enter in the window of perturbation background.
The resulting equations, after implication of perturbation strategy, are then used
to construct $f(R,T,Q)$ collapse equation. Then, we have used well-known Harrison-Wheeler state equation that has related peculiarly
the profiles of energy density and pressure components via stiffness of fluid
content. After considering a viable configurations of $f(R,T,Q)$
model, we have examined its impact in the definitions of modified hydrodynamical equation. The corresponding
constraints at both N and pN are evaluated. We observed that extra degrees of freedom induced from
 $f(R,T,Q)$ gravity try to produce
obstacles in the evolutionary phases of anisotropic compact star, thereby pushes the system to enter in unstable window.

Chandrasekhar \cite{6} calculated a specific value of the adiabatic index, i.e., $4/3$, for the
stability regime of the locally isotropic spherical relativistic objects. After this,
many astrophysicists investigated the these regions by taking
various choices of matter as well as geometric configurations. We have pointed out a key role of
stiffness parameter in the maintenance of stable backgrounds. We explore that the adiabatic index
have the influence of extra curvature ingredients due to matter curvature coupling
in static background. It is found the self-gravitating celestial object remains in stable state until
it satisfies (\ref{35}) and (\ref{36}) for N and
PN regimes, respectively. Once, the system fail to comply with the
prescribed ranges, it will enter into the unstable regime. We conclude that the
extra curvature terms due to $f(R,T,Q)$ theory makes the system more stable with the evolution of time,
thereby slowing down the collapse rate. It is noted that
with the zero existence of non-minimal coupling of matter and geometry,
these outcomes supports the results obtained in $f(R,T)$ findings \cite{z1}.

\vspace{0.3cm}

\section*{Acknowledgment}

This work was partially supported by the JSPS KAKENHI
Grant Number JP 25800136 and the research-funds presented by
Fukushima University (K.B.).

\renewcommand{\theequation}{A\arabic{equation}}
\setcounter{equation}{0}
\section*{Appendix A}

The quantities $\chi_i$'s of field equations are
\begin{align}\nonumber
\chi_1&=1+f_T-\frac{3R}{2}f_Q-\frac{\dot{A}\dot{f_Q}}{2A^3}+\frac{4\dot{A}^2}{A^4}f_Q-\frac{7\dot{A}}{2A^3}\dot{f_Q}
+\frac{2A'f'_Q}{AB^2}+\frac{A'^2f_Q}{A^2B^2}+\frac{A''f_Q}{AB^2}\\\nonumber
&-\frac{\dot{B}\dot{f_Q}}{2A^2B}-\frac{\dot{A}\dot{B}}{A^3B}f_Q-\frac{B'f'_Q}{2B^3}-\frac{A'B'}{AB^3}f_Q
-\frac{\dot{C}\dot{f_Q}}{A^2C}-\frac{2\dot{A}\dot{C}}{A^3C}f_Q+\frac{C'f'_Q}{CB^2}+\frac{2A'C'}{AB^2C}f_Q,\\\nonumber
\chi_2&=-\frac{11}{2A^3}f_Q\dot{A}-\frac{f_Q\dot{B}}{2A^2B}-\frac{f_Q\dot{C}}{A^2C},~
\chi_3=\frac{f_Q}{CB^2}C'+\frac{f'_Q}{B^2}-\frac{2A'f_Q}{AB^2}-\frac{B'f_Q}{2B^3},\\\nonumber
\chi_4&=\frac{f''_Q}{2B^2}+\frac{4B'^2f_Q}{B^4}-\frac{B''}{B^3}f_Q-\frac{5B'f'_Q}{2B^3}-\frac{\dot{B}\dot{f_Q}}{2A^2B}
+\frac{\dot{B}^2f_Q}{A^2B^2},\\\nonumber
\chi_5&=\frac{3\dot{C}^2f_Q}{A^2C^2}-\frac{3\dot{C}\dot{f_Q}}{2A^2C}-\frac{3C'f'_Q}{2B^2C}+\frac{3C'^2f_Q}{B^2C^2},\\\nonumber
\chi_6&=1+f_T-\frac{3}{2}Rf_Q-\frac{3B'^2f_Q}{B^4}+\frac{2B''}{B^3}f_Q+\frac{B'}{B^3}f'_Q-\frac{5\dot{B}\dot{f_Q}}{2A^2B}
-\frac{3\dot{B}^2f_Q}{A^2B^2}\\\nonumber
&-\frac{B\ddot{B}}{A^2B^2}f_Q-\frac{\ddot{f_Q}}{2A^2}+\frac{\dot{A}\dot{f_Q}}{2A^3}+\frac{\dot{A}\dot{B}}{A^3B}f_Q+\frac{A'f'_Q}{2AB^2},\\\nonumber
\chi_7&=\frac{4B'}{B^3}f_Q+\frac{A'}{2AB^2}f_Q,~\chi_8=\frac{\dot{B}}{2A^2B}f_Q+\frac{\dot{A}f_Q}{2A^3}-\frac{5\dot{B}f_Q}{2A^2B},\\\nonumber
\chi_9&=\frac{\dot{C}\dot{f_R}}{A^2C}-\frac{2\dot{C}}{A^2C^2}f_Q,~\chi_{11}=\frac{5\dot{A}f_Q}{2A^3}-\frac{\dot{f_Q}}{A^2},\\\nonumber
\chi_{10}&=\frac{5\dot{A}\dot{f_Q}}{2A^3}-4\frac{\dot{A}^2}{A^4}f_Q+\frac{\ddot{A}}{A^3}f_Q-\frac{\ddot{f_Q}}{2A^2}+\frac{A'f'_Q}{2AB^2}-\frac{A'^2f_Q}{A^2B^2},\\\nonumber
\chi_{12}&=1+f_T-\frac{3}{2}Rf_Q-\frac{5\dot{C}^2}{A^2C^2}f_Q+\frac{2C'}{CB^2}f_Q-\frac{2\dot{C}\dot{f_Q}}{A^2C}-\frac{\ddot{C}f_Q}{A^2C}
-\frac{\ddot{f_Q}}{2A^2}+\frac{\dot{A}\dot{f_Q}}{2A^3}\\\nonumber
&+\frac{\dot{A}\dot{C}}{CA^3}f_Q+\frac{A'f'_Q}{2AB^2}+\frac{A'C'}{ACB^2}f_Q+\frac{2C'f'_Q}{B^2C}+\frac{C'^2f_Q}{B^2C^2}+\frac{C''}{B^2C}f_Q
-\frac{\dot{B}\dot{f_Q}}{2A^2B}\\\nonumber
&+\frac{f''_Q}{2B^2}-\frac{\dot{C}\dot{B}}{CA^2B}f_Q-\frac{B'f'_Q}{2B^3}-\frac{C'B'}{CB^3}f_Q,\\\nonumber
\chi_{13}&=\frac{\dot{C}\dot{A}}{2A^3C^2}f_Q-\frac{2\dot{C}}{A^2C}f_Q-\frac{\dot{f_Q}}{A^2}-\frac{\dot{B}f_Q}{2A^2B},\\\nonumber
\chi_{14}&=\frac{2C'}{B^2C}f_Q+\frac{A'f_Q}{2AB^2}+\frac{2C'f_Q}{B^2C}+\frac{f'_Q}{B^2}-\frac{B'f_Q}{2B^3},\\\nonumber
\chi_{15}&=\frac{5B'f'_Q}{2B^3}-\frac{4B'^2}{B^4}f_Q+\frac{B''}{B^3}f_Q-\frac{f''_Q}{2B^2}+\frac{\dot{B}\dot{f_Q}}{2A^2B}-\frac{\dot{B}^2f_Q}{A^2B^2},\\\nonumber
\chi_{16}&=\frac{5\dot{A}}{2A^3}\dot{f_Q}-\frac{4\dot{A}^2}{A^4}f_Q+\frac{\ddot{A}}{A^3}f_Q-\frac{\ddot{f_Q}}{2A^2}-\frac{A'^2f_Q}{A^2B^2}+\frac{A'f'_Q}{2AB^2}.
\end{align}

\vspace{0.25cm}

\begin{thebibliography}{40}

\bibitem{zs1} S. Weinberg, Rev. Mod. Phys. \textbf{61}, 1 (1989);
P. J. E. Peebles and B. Ratra, Rev. Mod. Phys. \textbf{75}, 559 (2003);
V. Husain and B. Qureshi, Phys. Rev. Lett. \textbf{116}, 061302 (2016).

\bibitem{zs2} D. Pietrobon, A. Balbi, and D. Marinucci, Phys. Rev. D \textbf{74},
043524 (2006); T. Giannantonio et al., Phys. Rev. D \textbf{74}, 063520 (2006); A. G. Riess et al.,
Astrophys. J. \textbf{659}, 98 (2007).

\bibitem{zs3} A. Qadir, H. W. Lee, and K. Y. Kim, Int. J. Mod. Phys. D \textbf{26}, 1741001 (2017).

\bibitem{ya3} S. Nojiri and S. D. Odintsov, eConf C {\bf 0602061}, 06 (2006) [Int. J. Geom. Meth. Mod. Phys.
\textbf{4}, (2007) 115] [hep-th/0601213].

\bibitem{v41} T. P. Sotirou and V. Faraoni, Rev. Mod. Phys. \textbf{82}, 451 (2010).

\bibitem{b2b} S. Nojiri and S. D. Odintsov, Phys. Rep. \textbf{505}, 59 (2011).

\bibitem{b2a} S. Capozziello and M. D. Laurentis, Phys. Rep. \textbf{509}, 167 (2011).

\bibitem{R-DE-MG}
%
S.~Capozziello and V.~Faraoni, \textit{Beyond Einstein Gravity}
(Springer, Dordrecht, 2010);\
%
  K.~Bamba, S.~Capozziello, S.~Nojiri and S.~D.~Odintsov,
  Astrophys.\ Space Sci.\  {\bf 342}, 155 (2012);\
%
  A.~de la Cruz-Dombriz and D.~S\'{a}ez-G\'{o}mez,
  Entropy {\bf 14}, 1717 (2012);\
%
  A.~Joyce, B.~Jain, J.~Khoury and M.~Trodden,
  Phys.\ Rept.\  {\bf 568}, 1 (2015);
%
  K.~Koyama,
  arXiv:1504.04623 [astro-ph.CO];\
%
  K.~Bamba, S.~Nojiri and S.~D.~Odintsov,
  arXiv:1302.4831 [gr-qc];\
%
   K.~Bamba and S.~D.~Odintsov,
   arXiv:1402.7114 [hep-th];\
%
  Symmetry {\bf 7}, 220 (2015)
  [arXiv:1503.00442 [hep-th]];
  Z.~Yousaf,~K.~Bamba~and~M.~Z.~Bhatti,~Phys. Rev. D \textbf{95}, 024024 (2017) [arXiv:1701.03067 [gr-qc]].

\bibitem{no1} S. Nojiri, and S. D. Odintsov, Phys. Rev D \textbf{68}, 123512 (2003).

\bibitem{z1fr} M. Sharif and Z. Yousaf, Gen. Relativ. Gravit. \textbf{47}, 48 (2015);
M. Sharif and Z. Yousaf, Can. J. Phys. \textbf{93}, 905 (2015); M.
Z. Bhatti and Z. Yousaf, Int. J. Mod. Phys. D \textbf{26}, 1750029
(2017); ibid. Int. J. Mod. Phys. D \textbf{26}, 1750045 (2017); M.
Z. Bhatti and Z. Yousaf, Eur. Phys. J. C \textbf{76}, 219 (2016)
[arXiv1604.01395 [gr-qc]].

\bibitem{kkk1} Z. Yousaf, Eur. Phys. J. Plus \textbf{132}, 71 (2017).

\bibitem{z2frt} T. Harko, F. S. N. Lobo, S. Nojiri and S. D. Odintsov,
Phys. Rev. D \textbf{84}, 024020 (2011); Z. Yousaf and M. Z. Bhatti, Eur. Phys. J. C \textbf{76}, 267
(2016) [arXiv:1604.06271 [physics.gen-ph]];
M. Sharif~and~Z. Yousaf,~Astrophys.~Space~Sci.~\textbf{354},~471~(2014); Yousaf, Z., ~Bamba,
K.~and~Bhatti,~M.~Z.:~Phys. Rev. D \textbf{95}, 024024 (2017)
[arXiv:1701.03067 [gr-qc]].

\bibitem{g11} S. Nojiri and S. D. Odintsov, Phys. Lett. B \textbf{631}, 1 (2005).

\bibitem{z3frtrmn} S. D. Odintsov and D. S\'{a}ez-G\'{o}mez, Phys. Lett. B, \textbf{725}, 437 (2013).

\bibitem{z3frtrmn1} Z. Haghani, T. Harko, F. S. N. Lobo, H. R. Sepangi and S. Shahidi, Phys. Rev. D, \textbf{88}, 044023 (2013);
Z. Yousaf, M. Z. Bhatti and U. Farwa, Class. Quantum Grav. \textbf{34}, 145002 (2017).

\bibitem{6} S. Chandrasekhar, Astrophys. J. \textbf{ 140}, 417 (1964).

\bibitem{ck1} L. Herrera, N. O. Santos and G. Le Denmat, Mon. Not. R. Astron. Soc. \textbf{237}, 257 (1989).

\bibitem{ck2} R. Chan, L. Herrera and N. O. Santos, Mon. Not. R. Astron. Soc. \textbf{265}, 533 (1993); ibid. \textbf{267}, 637 (1994).

\bibitem{ck3} R. Chan, Mon. Not. R. Astron. Soc. \textbf{316}, 588 (2000).

\bibitem{8} L. Herrera, G. Le Denmat and N. O. Santos, Gen. Relativ. Gravit. \textbf{44}, 1144 (2012).

\bibitem{10} Z. Haghani, T. Harko, F. S. N. Lobo, H. R. Sepangi and S. Shahidi, Phys. Rev. D \textbf{88}, 044023 (2013).

\bibitem{11} I. Ayuso et al., Phys. Rev. D \textbf{91}, 104003 (2015).

\bibitem{d11} E. Elizalde and S. I. Vacaru, Gen. Relativ. Gravit. \textbf{47}, 64 (2015)

\bibitem{far1} Z. Yousaf, M. Z. Bhatti and U. Farwa, Mon. Not. R. Astron. Soc.
\textbf{464}, 4509 (2017); ibid.
Eur. Phys. J. C \textbf{77}, 359 (2017) [arXiv:1705.06975 [physics.gen-ph]].

\bibitem{gc1} R. Garattini, J. Phys. Conf. Ser. \textbf{174}, 012066
(2009).

\bibitem{gc2} A. H. Ziaie, K. Atazadeh and S. M. M. Rasouli, Gen. Relativ.
Gravit. \textbf{43}, 2943 (2011).

\bibitem{gc3} S. G. Ghosh and S. D. Maharaj, Phys. Rev. D \textbf{85},
124064 (2012).

\bibitem{gc4} B. M. N\'{u}\~{n}ez, J. A. R. Cembranos, and \'{A}. de la
Cruz-Dombriz, AIP Conf. Proc. \textbf{1458}, 491 (2012),
[arXiv:1210.7968 [gr-qc]].

\bibitem{gc5} L. Reverberi, Phys. Rev. D \textbf{87}, 084005
(2013); Z. Yousaf, Eur. Phys. J. Plus \textbf{132}, 71 (2017); Eur.
Phys. J. Plus \textbf{132}, 276 (2017).

\bibitem{gc6} E. V. Arbuzova, A. D. Dolgov and L. Reverberi, Astropart. Phys.
\textbf{54}, 44 (2014).

\bibitem{gc6a} M. Sharif and Z. Yousaf, Eur. Phys. J. C \textbf{75}, 58 (2015).

\bibitem{gc7} M. Z. Bhatti, Eur. Phys. J. Plus \textbf{131}, 428 (2016);
M. Sharif and Z. Yousaf, Eur. Phys. J. C  \textbf{75}, 194 (2015) [arXiv:1504.04367v1 [gr-qc]].

\bibitem{gc8} M. Fathi and M. Mohseni, Eur. Phys. J. Plus \textbf{131}, 360
(2016).

\bibitem{gc9} Z. Yousaf and M. Z. Bhatti, Mon. Not. R. Astron. Soc. \textbf{458}, 1785
(2016).

\bibitem{gc10} S. Chakrabarti and N. Banerjee, Eur. Phys. J. Plus \textbf{131}, 144
(2016).

\bibitem{gc11} M. Sharif and Z. Yousaf, Int. J. Theor. Phys. \textbf{55}, 470
(2016)

\bibitem{gc12} J. -Q. Guo and P. S. Joshi, Phys. Rev. D \textbf{94}, 044063 (2016).

\bibitem{gc13} Z. Yousaf, K. Bamba and M. Z. Bhatti, Phys. Rev. D \textbf{93}, 124048 (2016)
[arXiv1606.00147 [gr-qc]].

\bibitem{gc14} C. -Y. Zhang, Z. -Y. Tang and B. Wang, Phys. Rev. D \textbf{94}, 104013 (2016)
arXiv:1608.04836 [gr-qc].

\bibitem{gc15} S. Satin, D. Malafarina and P. S. Joshi, Int. J. Mod. Phys. D
\textbf{25}, 1650023 (2016).

\bibitem{gc16} G. Abbas and M. Tahir, Eur. Phys. J. C \textbf{77}, 537 (2017) [arXiv:1707.08472 [gr-qc]].

\bibitem{ya12} K. S. Adhav, Astrophys. Space Sci. \textbf{339}, 365 (2012).

\bibitem{sahoo1} S. K. Sahu, S. K. Tripathy, P. K. Sahoo and A. Nath, Chin. J. Phys. \textbf{55}, 862 (2017);
P. K. Sahoo, P. Sahoo, B. K. Bishi and S. Ayg\"{u}n, Mod. Phys. Lett. \textbf{32}, 1750105 (2017).

\bibitem{herr1} L. Herrera, A. Di Prisco, J.L. Hernández-Pastora, and N.O. Santos. Phys. Lett. A,
237, 113 (1998); L. Herrera. Int. J. Mod. Phys. D, 20, 1689 (2011); L. Herrera and N.O. Santos. Phys. Rev. D, 70, 084004 (2004); L. Herrera, A. Di Prisco, and W. Barreto. Phys. Rev. D, 73, 024008 (2006); L. Herrera. Int. J. Mod. Phys. D, 15, 2197 (2006).

\bibitem{zs25} S. Capozziello, M. De Laurentis, S. D. Odintsov, and
A. Stabile, Phys. Rev. D \textbf{83}, 064004 (2011);
S. Capozziello, M. De Laurentis, I. De Martino, M. Formisano, and S. D. Odintsov, Phys. Rev. D \textbf{85}, 044022 (2012).

\bibitem{zs26} J. A. R. Cembranos, \'{A}. de la Cruz-Dombriz and B. M. N\'{u}\~{n}ez,
J. Cosmol. Astropart. Phys. \textbf{04}, 021 (2012).

\bibitem{zs31} M. Sharif and Z. Yousaf, Eur. Phys. J. C  \textbf{75}, 194 (2015) [arXiv:1504.04367v1 [gr-qc]];
M. Sharif and Z. Yousaf, Astrophys. Space Sci. \textbf{357}, 49
(2015); ibid. \textbf{351}, 351 (2014); M. Z. Bhatti and Z. Yousaf,
Eur. Phys. J. C \textbf{76}, 219 (2016) [arXiv1604.01395 [gr-qc]];
M. Z. Bhatti and Z. Yousaf, Ann. Phys. \textbf{387}, 253 (2017);
M. Z. Bhatti, Z. Yousaf and S. Hanif,
Phys. Dark Universe \textbf{16}, 34 (2017).

\bibitem{13a} Z. Yousaf, K. Bamba and M. Z. Bhatti, Phys. Rev. D \textbf{93},
064059 (2016) [arXiv1603.03175 [gr-qc]].

\bibitem{13b} E. H. Baffou, M. J. S. Houndjo and J. Tossa, Astrophys. Space Sci. \textbf{361}, 376 (2016) [arXiv:1608.03273[gr-qc]].

\bibitem{14} T. Harko, F. S. N. Lobo, S. Nojiri and S. D. Odintsov, Phys. Rev. D \textbf{84}, 024020 (2011).

\bibitem{15} C. Misner and D. Sharp, Phys. Rev. D \textbf{136}, B571 (1964).

\bibitem{16} B. K. Harrison, K. S. Throne, M. Wakano and J. A. Wheeler, {\it Gravitation Theory and Gravitational Collapse} (University of Chicago press, 1965).

\bibitem{eos1} Z. Yousaf, M. Sharif, M. Ilyas and M. Z. Bhatti, Eur. Phys. J. C \textbf{77}, 691 (2017) [arXiv:1710.05717 [gr-qc]];
M. Z. Bhatti, M. Sharif, Z. Yousaf and M. Ilyas, Int. J. Mod. Phys. D \textbf{27}, 1850044 (2018);
Z. Yousaf, M. Z. Bhatti and M. Ilyas, Eur. Phys. J. C 78, 307 (2018) [arXiv:1804.04953 [physics.gen-ph]].

\bibitem{z1} Z. Yousaf and M. Z. Bhatti, Eur. Phys. J. C Eur. Phys. J. C \textbf{76}, 267 (2016) [arXiv1604.06271 [physics.gen-ph]].

\bibitem{z2} M. Sharif and Z. Yousaf, Astrophys. Space Sci. \textbf{355},
317 (2015).

\bibitem{zzz3} C. M. Will, \emph{Theory and Experiment in Gravitational Physics} (Cambridge Univ.
Press, 1993); N. Straumann, \emph{General Relativity: with Aplication to
Astrophysics} (Springer, 2004).

\bibitem{zzz2} M. Sharif and Z. Yousaf, Astropart. Phys. \textbf{56}, 19 (2014).

\end{thebibliography}
\end{document}